### 23.1 T-REX: A 68-to-567μs/Token 0.41-to-3.95μJ/Token Transformer Accelerator with Reduced External Memory Access and Enhanced Hardware Utilization in 16nm FinFET

Seunghyun Moon[1], Mao Li[1], Gregory K. Chen[2], Phil C. Knag[2],
Ram Kumar Krishnamurthy[2], Mingoo Seok[1]

[1]Columbia University, New York, NY
[2]Intel, Hillsboro, OR

Transformer, a recent mainstream model in deep learning, has revolutionized a wide range of AI applications, which motivates a surge in research to develop energy-efficient hardware accelerators. Most prior efforts have concentrated on enhancing on-chip computational energy efficiency through several strategies such as encoder-only models [1-7], quantization/sparsity [8-18], and layer pruning [19]. However, recent works [20,21] show that external memory access (EMA) dominates total energy consumption. Our analysis based on [22,23] also indicates that EMA accounts for up to 81% of the total energy usage (Fig. 23.1.1). Additionally, we recognize that the prior works exhibit low hardware utilization, as low as 9% in [4], which negatively impacts latency performance.

In light of this, we present a novel transformer accelerator named T-REX to address the challenges of EMA and hardware utilization. To reduce EMA, based on [24], we developed a factorizing training model that decomposes each weight matrix into a dense matrix shared across all layers ($W_S$) and a highly sparse matrix distinct to each layer ($W_D$). During runtime, T-REX needs to preload $W_S$ only once, significantly reducing EMA. To further scale down EMA, we compress $W_S$ and $W_D$ using several advanced compression techniques. Next, we propose a dynamic batching technique, where T-REX monitors input lengths and, if the input is 2× (4×) smaller than the maximum input length of T-REX, it processes 2 (4) inputs simultaneously by reconfiguring its dataflow. This approach reduces EMA by minimizing the number of parameter loads and also enhances hardware utilization. Finally, we developed two-direction accessible register files (TRFs) within the computing cores to load and store a matrix in both row-by-row (R-R) and column-by-column (C-C) fashions. They eliminate the latency overhead caused by accessing SRAMs multiple times, additionally enhancing hardware utilization. Combining the proposed techniques, we prototyped the T-REX test chip in 16nm FinFET. Measurement results show that T-REX can reduce EMA by 31-65.9× and improve hardware utilization by 1.2-3.4× across four well-known transformer workloads [25-28]. It achieves 68-567μs/token and 0.41-3.95μJ/token, including EMA.

Figure 23.1.2 shows the microarchitecture of T-REX, designed for energy-efficient and low-latency inference with factorized and compressed transformer models. It consists of an IO interface, a RISC-V core-based top controller, a DMA, a global buffer (GB), four dense matrix-multiplication (DMM) cores, four sparse matrix-multiplication (SMM) cores, and two auxiliary function units (AFUs). The GB stores compressed $W_S$, compressed $W_D$ for one layer, and intermediate data. Each DMM core includes a lookup table (LUT)-based non-uniform dequantizer, input and output buffers, an accumulator, and 4×4 processing elements (PEs). Each PE contains 4×4 multiply-and-accumulate (MAC) units, each of which has a 4b multiplier and a 32b accumulator and performs a 16b (8b, 4b) MAC operation over 16 (4, 1) cycles. Each PE is implemented to perform a 4×4 outer product, allowing DMM cores with 4×4 PEs to generate 16×16 output elements simultaneously and compute tiled matrix multiplication (MM) with a tile size of 16×16. On the other hand, each SMM core consists of a uniform dequantizer, input and output buffers, a sparse line buffer, a bias buffer, an accumulator, and 8×8 MAC units. The MAC units are identical to those in the DMM cores. The SMM cores can be configured to perform row (column) products depending on which input matrix is sparse. Non-zero elements (NZs) in the sparse matrix are loaded into the line buffer, while the corresponding rows (columns) are loaded into the input buffer. This sparsity-aware switching feature enables more efficient computation. Finally, the AFU includes input and output buffers, two LUTs for the exponential and GELU functions, 64 integer arithmetic units (IAUs), 16 floating-point arithmetic units (FAUs), BF16↔INT32 converters. The AFUs perform softmax, layer normalization, GELU, and residual connection. For example, in the softmax, the AFU utilizes the LUT for the exponential function and then uses the IAUs to evaluate the remaining computations. Depending on the transformer model, the converters and the FPUs can be used for higher accuracy requirements.

Figure 23.1.3 shows our training model, which replaces a weight matrix W with the product of two submatrices: $W=W_S \cdot W_D$. During runtime, $W_S$ is loaded only once, which substantially reduces EMA. Additionally, $W_D$ is trained to be sparse by adding a regularization term to the loss function, ensuring that each column contains a fixed number of NZs. As $W_D$ becomes highly sparse, we store only the indices and values of the NZs. This compressed format, although similar to compressed sparse column format, does not require storing the column pointer, enabling additional EMA reduction. The proposed training model reduces EMA by 8.5-10.7× across four transformer workloads. The main operation of T-REX is sequential MM, $(X \cdot W_S) \cdot W_D$, where X is the input matrix. We choose this computing order over $X \cdot (W_S \cdot W_D)$ because the hidden size of $W_S$ is much smaller than that of $W_S \cdot W_D$, reducing the total number of MACs. Furthermore, even compared to X·W, the chosen computation requires 1-2.14× fewer MAC operations across the tested models.

To further reduce EMA, we apply 16b-to-4b non-uniform quantization to $W_S$, reducing the size of $W_S$ by 4× with negligible accuracy loss. We also apply 8b-to-5b delta encoding (i.e., storing the difference of two consecutive values) to indices of $W_D$. Smaller delta values allow us to use narrower bitwidth, improving the compression ratio. To minimize the delta values without changing $W_S \cdot W_D$, we rearranged the columns of $W_S$ and the corresponding rows of $W_D$. We also apply 16b-to-6b uniform quantization to values of $W_D$. To improve the compression ratio, we normalize each value of $W_D$ with a layer-specific scale (M-m) and offset (m), making the distribution symmetric around zero and maximizing the available range and precision of the uniform quantization. The proposed compression techniques enable an additional EMA reduction of 2.1-2.9× across the target models.

Figure 23.1.3 bottom illustrates the hardware support for the main computations in T-REX. The DMM cores handle the first part of the main computation, i.e., $X \cdot W_S$. The input data and $W_S$ are loaded, and the LUT-based non-uniform dequantizer decompresses the 4b non-uniformly quantized $W_S$ to 16b integers, followed by MM within the PEs. For the encoder and decoder layers, as well as for the attention and feed-forward layers, we define separate $W_S$ and maintain independent quantized values. The LUT is reconfigured to accommodate these different quantization settings. Next, the SMM cores perform the second MM, i.e., $(X \cdot W_S) \cdot W_D$. To load the input, delta-encoded indices are used for addressing. Instead of explicit decoding, we use relative addressing to load the corresponding columns of the input matrix. For values of $W_D$, the uniform dequantizer restores the 6b values of $W_D$ back to 16b using the stored scale and offset. The MAC units then perform the MM, considering only NZs.

We developed a dynamic batching technique to further reduce EMA and improve hardware utilization. T-REX supports the maximum input length of 128. If the input length is between 128 and 65, we configure the dataflow to take one input and produce one output (Fig. 23.1.4 top left). On the other hand, if the input length is between 64 and 33 (32 or less), as shown in Fig. 23.1.4 top right (bottom left) we reconfigure the dataflow to process two (four) inputs simultaneously by specifying which submatrices the DMM/SMM cores use, and which blocks are utilized inside the AFUs. Note that data movement between computing blocks occurs via memory operations, rather than through dedicated buses. Therefore, it incurs <0.1% area overhead to support the dataflow reconfiguration. The proposed dynamic batching technique is particularly effective when the model processes many inputs with short lengths, such as in BERT-Large. It reduces EMA by allowing T-REX to reuse parameters across multiple inputs and improves hardware utilization by up to 3.31×, leading to reduced latency.

Figure 23.1.5 shows a complexity associated with MMs where matrices need to be accessed in different directions. In DMMs using an outer product, X ($W_S$) needs to be loaded C-C (R-R), and the result Y needs to be stored C-C for the subsequent column product in SMMs. The SMM output Z also needs to be stored in the appropriate direction depending on the next operation; here, it is assumed to be stored R-R. However, if all buffers allow only R-R access as in the conventional memory architecture, it results in wasted clock cycles due to the significant number of SRAM accesses. To address this, we implemented TRFs as the input and output buffers, which contain square-shaped submatrices and allow data access in both row and column directions. These TRF-based buffers eliminate the waste of SRAM access that would otherwise cause all PEs to be idle, thereby improving hardware utilization by 12-20%.

We prototyped the T-REX test chip in 16nm FinFET with a total area of 10.15mm² (Fig. 23.1.7). The measurement results show that T-REX operates at 60-450MHz across 0.45 to 0.85V, consuming 7.12 to 152.5mW. Figure 23.1.6 shows the four transformer models that we trained. The proposed training and compression techniques reduce the parameter size by 15.9-25.5× with minimal accuracy loss. When performing inference with these models, T-REX requires 31 to 65.9× less EMA and exhibits 1.2 to 3.4× higher hardware utilization. We compared T-REX with the previous accelerators. For those works that do not consider EMA, we estimated the energy cost at 3.7pJ/b and the latency cost at 6.4GB/s, both based on the LPDDR3 SDRAM [22,23]. T-REX achieves 68 to 567μs/token and 0.41 to 3.95μJ/token, marking significant improvements across several workloads over prior works.

*Acknowledgement:*
This work was supported in part by an SRC AIHW program (Task 3160.002) and by COGNISENSE, one of seven centers in JUMP 2.0, an SRC program sponsored by DARPA.

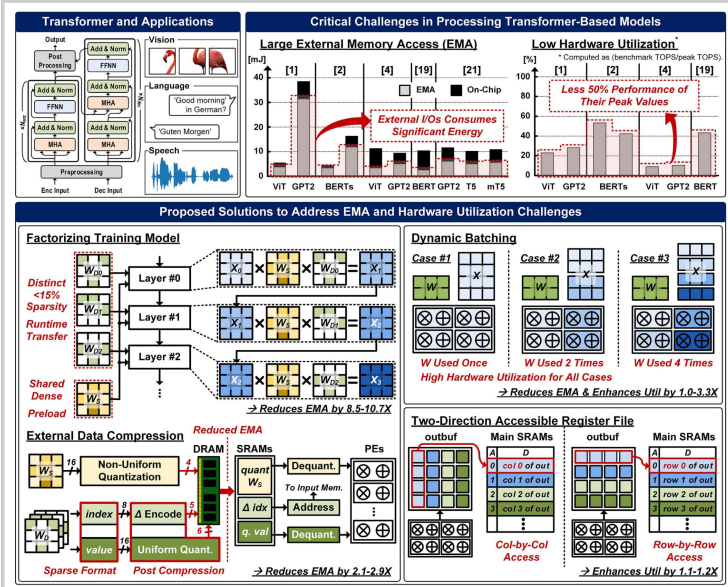

Figure 23.1.1: Challenges in transformer processing and proposed solutions.

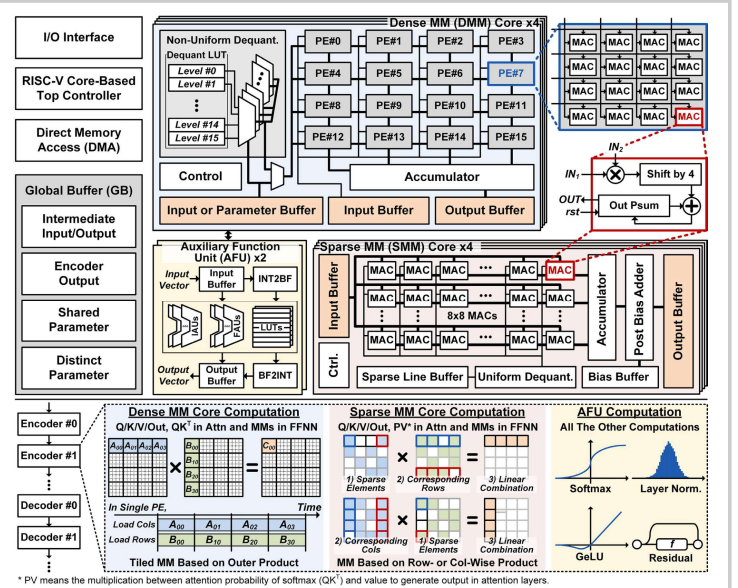

Figure 23.1.2: Overall architecture of T-REX.

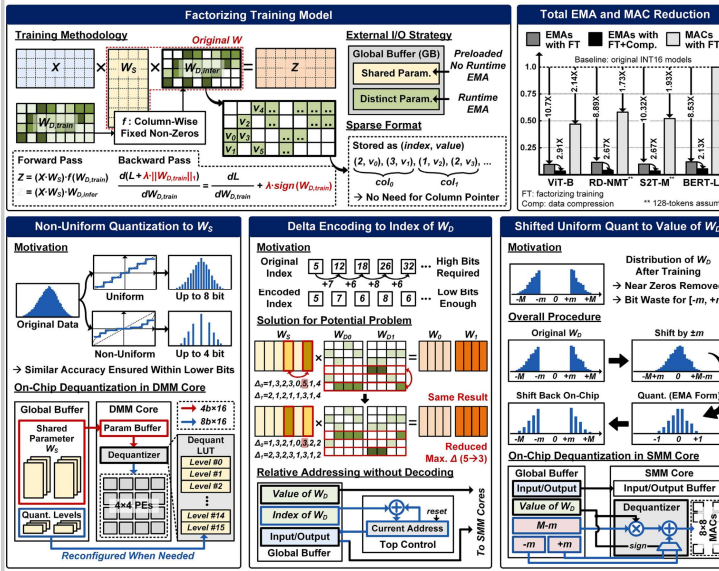

Figure 23.1.3: Factorizing training and compressions with hardware support.

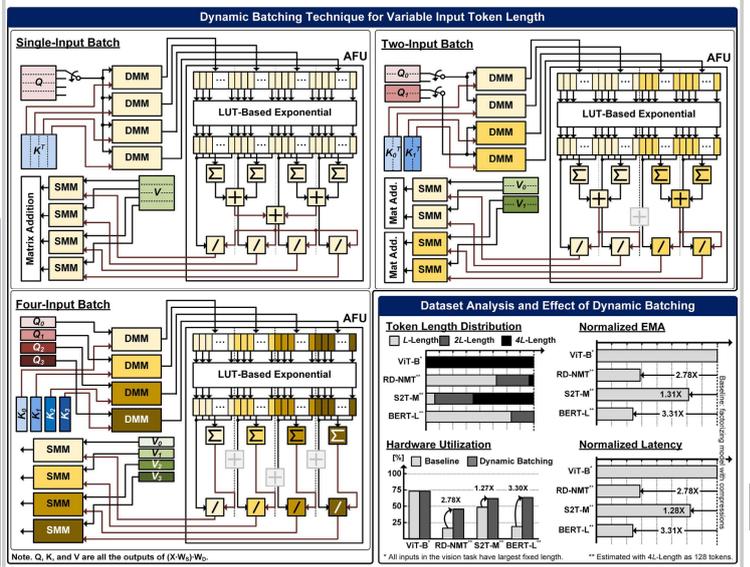

Figure 23.1.4: Dynamic batching technique for variable input token length.

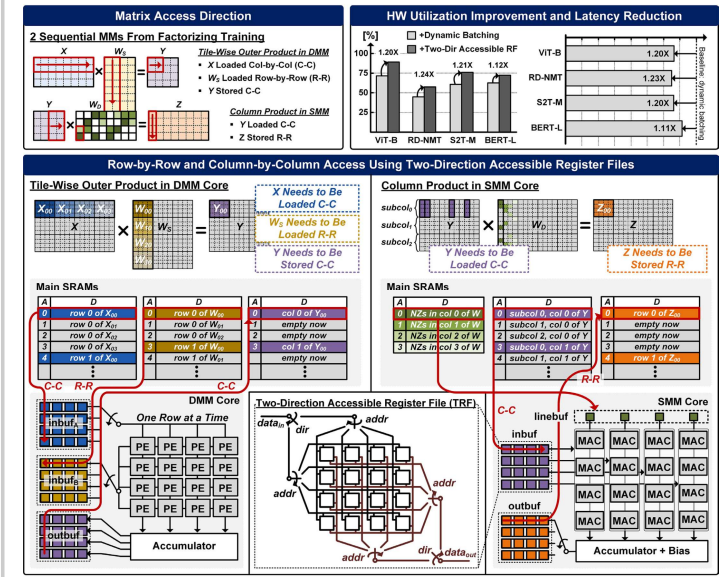

Figure 23.1.5: Input and output buffers based on two-direction accessible register file.

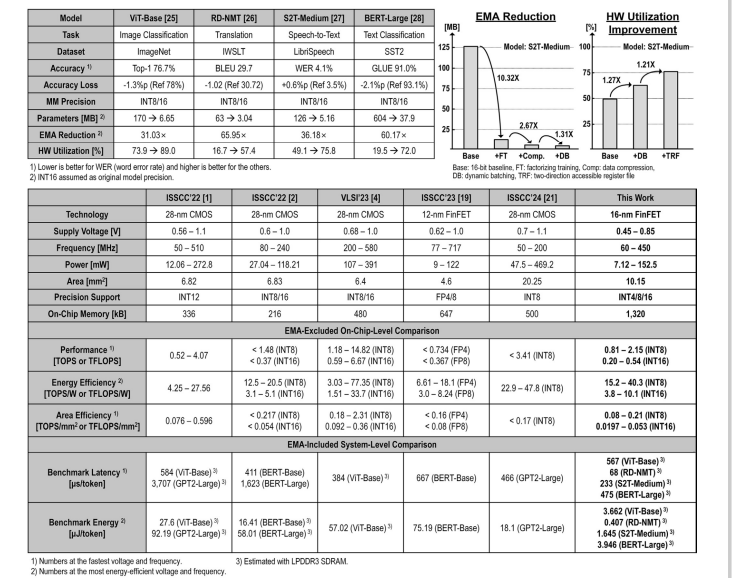

Figure 23.1.6: Measurement result and comparison table.

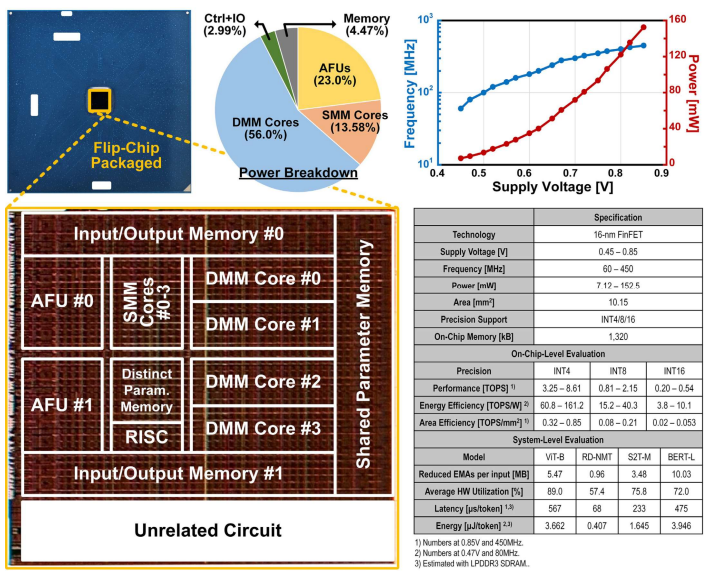

Figure 23.1.7: Chip photograph and performance summary.